\documentclass [prb,twocolumn, amsmath,amssymb]{revtex4}
\setcitestyle{numbers,square}
\usepackage{appendix}
\usepackage{array}
\usepackage{amsmath,amssymb}
\usepackage{tabularx}
\usepackage{graphicx}
\usepackage{bmpsize}
\usepackage{bm}
\usepackage{color}
\usepackage{amssymb}
\usepackage [version=3]{mhchem}
\usepackage{newtxmath}
\usepackage{hyperref}
\usepackage{url}
\usepackage{physics}
\usepackage{here}
\DeclareMathAlphabet{\mathpzc}{OT1}{pzc}{m}{it}

\begin{document}
\title{
Topological edge and corner states and fractional corner charges in blue phosphorene
}
\author{Tenta Tani}
\affiliation{Department of Physics, Osaka University, Toyonaka, Osaka 560-0043, Japan}
\author{Masaru Hitomi}
\affiliation{Department of Physics, Osaka University, Toyonaka, Osaka 560-0043, Japan}
\author{Takuto Kawakami}
\affiliation{Department of Physics, Osaka University, Toyonaka, Osaka 560-0043, Japan}
\author{Mikito Koshino}
\affiliation{Department of Physics, Osaka University, Toyonaka, Osaka 560-0043, Japan}
\date{\today}

\begin{abstract}
We theoretically study emergent edge and corner states in monolayer blue phosphorus (blue phosphorene) using the first-principles calculation and  tight-binding model. 
We show that the existence of the Wannier orbitals at every bond center yields edge states both in zigzag and armchair nanoribbons.
The properties of the edge states can be well described by a simple effective Hamiltonian for uncoupled edge orbitals,
where the structural relaxation near the boundary
significantly affects the edge band structure.
For corner states, we examine two types of corner structures
consisting of zigzag and armchair edges,
where we find that multiple corner states emerge in the bulk gap
as a consequence of hybridization of edge and corner uncoupled 
orbitals.
In the armchair corner, in particular,
we demonstrate that corner states appear right at the Fermi energy, 
which leads to the emergence of fractional corner charge
due to filling anomaly.
Finally, we discuss the relationship between blue phosphorene and black phosphorene, and show that 
two systems share the equivalent Wannier orbital positions and similar edge/corner state properties even though
their atomic structures are totally different.

\end{abstract}

\maketitle
\section{introduction}
\label{sec:intro}
Since the discovery of graphene, a variety of two-dimensional materials have attracted considerable attentions as the post-graphene semiconductors~\cite{Silicene_Fleurence2012,Silicene_Garcia2011,Silicene_Vogt2012,Silicene_Kharadi2020,Silicene_Molle2018,Germanene_Davila2014,Germanene_Yuhara2018,Germanene_Cahangirov2009,Germanene_Acun2015,hBN_Lin2012,hBN_Li2015,hBN_Gorbachev2011,BlackP_wei2014,Yu2015,Guo2014,Khandelwal2017,Hitomi2021,Islam2018,Fukuoka2015,BlackP_Ezawa2014,Tomanek2014,Zhang2016,Zhang2018,Liu2021,Zhuang2018,Lu2019,Zhou2021,Xie2014,Arcudia2020,Peng2016,BluePedge_Zhang2020}.
One of these is black phosphorene \cite{Yu2015,Guo2014,Khandelwal2017,Hitomi2021}, 
in which phosphorus atoms are arranged in a single layer of puckered honeycomb lattice. 
Although the low-energy band structure of graphene near the Fermi energy is described by $2p_z$ orbital only, that of black phosphorene consists of three orbitals of $3p_x$, $3p_y$ and $3p_z$. 
Due to this multi-orbital nature, black phosphorene hosts 
edge states both in zigzag and armchair boundaries \cite{BlackP_Ezawa2014,Islam2018,Fukuoka2015,Hitomi2021}
unlike graphene which has only zigzag edge states.


On the other hand, there is another two-dimensional allotrope called blue phosphorene, which is as stable as black phosphorene~\cite{Tomanek2014,Zhang2016,Zhang2018,Zhuang2018}.
Blue phosphorene also has a non-flat two-dimensional 
honeycomb lattice, but with a totally different atomic structure.
For instance, black phosphorene has four atoms in its unit cell due to its nonsymmorphic symmetry, while blue phosphorene hosts only two atoms in its unit cell.
It is an indirect semiconductor with a band gap of 2 eV \cite{Tomanek2014, Xie2014},
and expected as a promising candidate for optoelectronic devices.~\cite{Zhou2021, Zhuang2018, Zhang2016, Zhang2018}
Blue phosphorene has been recently synthesized \cite{Zhang2016},
and its physical properties were investigated experimentally~\cite{Zhuang2018, Zhang2018}.

The edge properties of blue phosphorene have also been addressed by previous theoretical works,
in terms of the electronic structures~\cite{Xie2014, Xiao2016}, passivation effect~\cite{Liu2015},
ferromagnetic effect~\cite{Hu2015} and electronic transport~\cite{An2018}.
In particular, it was shown that unpassivated zigzag and armchair ribbons host edge states~\cite{BluePedge_Zhang2020, Hu2015, Liu2015}.
Interestingly, the edge-state band structures in blue phosphorene resemble those in black phosphorene nanoribbons~\cite{Guo2014, Zhu2014}, even though
the two systems have totally different atomic structures and crystallographic symmetries.
In our previous work~\cite{Hitomi2021}, we showed that the topological origin of edge states in black phosphorene can be understood by using the center of the Wannier orbital,
which is a topological invariant.
The similarity in the edge-state natures between the two allotropes
suggests a certain topological relationship, which is not yet clear.

In this paper, 
we theoretically study edge and corner states in blue phosphorene using the first-principles calculation and tight-binding (TB) model,
and investigated their topological origins.
We show that zigzag and armchair edge states and the associated number of the edge bands
can be explained by considering the position of the Wannier orbitals,
in a similar manner to the black phosphorene~\cite{Hitomi2021}.
Similarity between blue and black phosphorenes
can be understood by the fact that
two systems can be deformed into a topologically equivalent model
through the deformation of the bond angles to $90^{\circ}$.
The properties of the edge states can be well described by a simple effective Hamiltoinian for uncoupled edge orbitals, where we find that the structural relaxation near the boundary
significantly modifies the edge band structure.

For corner states, we examine two types of corner structures
consisting of zigzag and armchair edges,
where we find that multiple corner states emerge in the bulk gap
as a consequence of hybridization of edge and corner uncoupled
orbitals.
These multiple corner states can also be well described by a simple effective Hamiltonian for uncoupled edge and corner orbitals.
In the armchair corner, in particular,
we demonstrate that corner states appear right at the Fermi energy, 
which leads to the emergence of fractional corner charge
due to filling anomaly~\cite{Benalcazar2019,Takahashi2021,Watanabe2020,watanabe2021fractional,Ren2021,Schindler2019,Hirayama2020,Wladimir2017,Kooi2021}.



This paper is organized as follows. 
In Sec.~\ref{sec:edge}, we perform density functional theory (DFT) calculations and construct the TB model. 
By using the TB model, we clarify the topological origin of the edge states. 
In Sec.~\ref{sec:corner}, we examine the edge and the corner states in two types of nanoflakes using effective edge-corner site model. 
We reveal that the isolated corner state emerges at the Fermi energy and a fractional corner charge is induced at the armchair-type corner.
In Sec.~\ref{sec_discussion}, we compare the result in blue phosphorene with that of black phosphorene.
Finally, we conclude this paper in Sec.~\ref{sec_conclusion}.

\section{Edge States of Blue Phosphorene}
\label{sec:edge}
\subsection{DFT calculations}
\label{subsec:DFT}
\begin{figure}
\begin{center}
   \includegraphics [width=85mm]{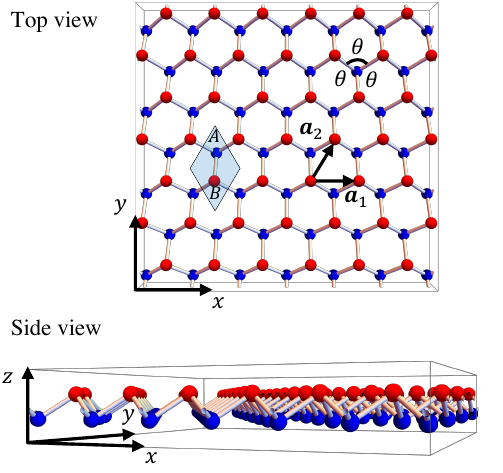}
   \caption{
   Atomic structure of blue phosphorene from the top and the side views. 
   All of the three bond angles are $\theta \sim 93.1^{\circ}$.
   The unit cell is shown as the blue rhombus, with A, B sublattices. 
   The red (blue) balls indicate atoms in the top (bottom) layer. 
            }\label{fig:lattice}
 \end{center}
 \end{figure}

Blue phosphorene has a buckled honeycomb structure shown in Fig.~\ref{fig:lattice}, where the A sublattice and B sublattice
are located on different two-dimensional planes.
The lattice constant is $a=3.28$ \AA \ and the buckling height (vertical distance between A and B sites) is $d=1.23$ \AA  ~\cite{Zhang2018,Zhang2016,Liu2021}. 
The structure belongs to the symmorphic group $P \bar{3} m 1$, which consists of rotoinversion $S_3=C_3 P$ along the $z$ axis, reflection about the $y-z$ plane and translations along the primitive lattice vectors $\vb*{a}_1$ and $\vb*{a}_2$. 
In this paper, we take the lattice vectors as shown in Fig.~\ref{fig:lattice}.

\begin{figure}
\begin{center}
   \includegraphics [width=85mm]{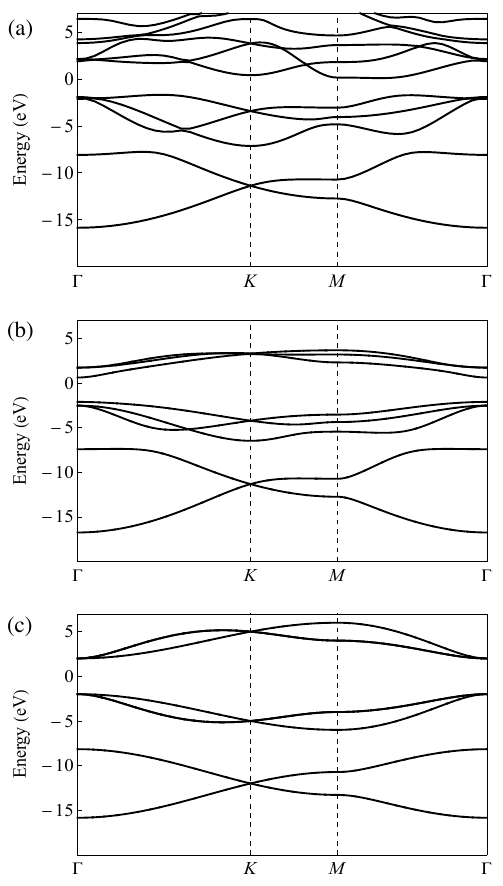}
   \caption{
   Electronic band structures of blue phosphorene obtained by (a) DFT calculation, (b) the Slater-Koster tight-binding model
   and (c) the $90^{\circ}$ model.
   For the DFT calculation, we employ the Vanderbilt ultrasoft pseudopotential with Perdew-Zunger exchange-correlation functional. 
   The cutoff energy of the plane-wave basis is 30 Ry, and the convergence criterion is $10^{-8}$ Ry in $12 \times 12 \times 1 \ \vb*{k}$-points mesh. The Fermi energy is set to $E = 0$.
            }\label{fig:bulkDFT90TBband}
 \end{center}
 \end{figure}

We perform first-principles band calculation by using the Quantum-ESPRESSO package \cite{QE2009,QE2017}.
Fig.~\ref{fig:bulkDFT90TBband}(a) shows the bulk band structure along the high-symmetry line of the first Brillouin zone, where we see that blue phosphorene is a semiconductor with a band gap of approximately $2$ eV. 
 
\begin{figure*}
\begin{center}
   \includegraphics [width=170mm]{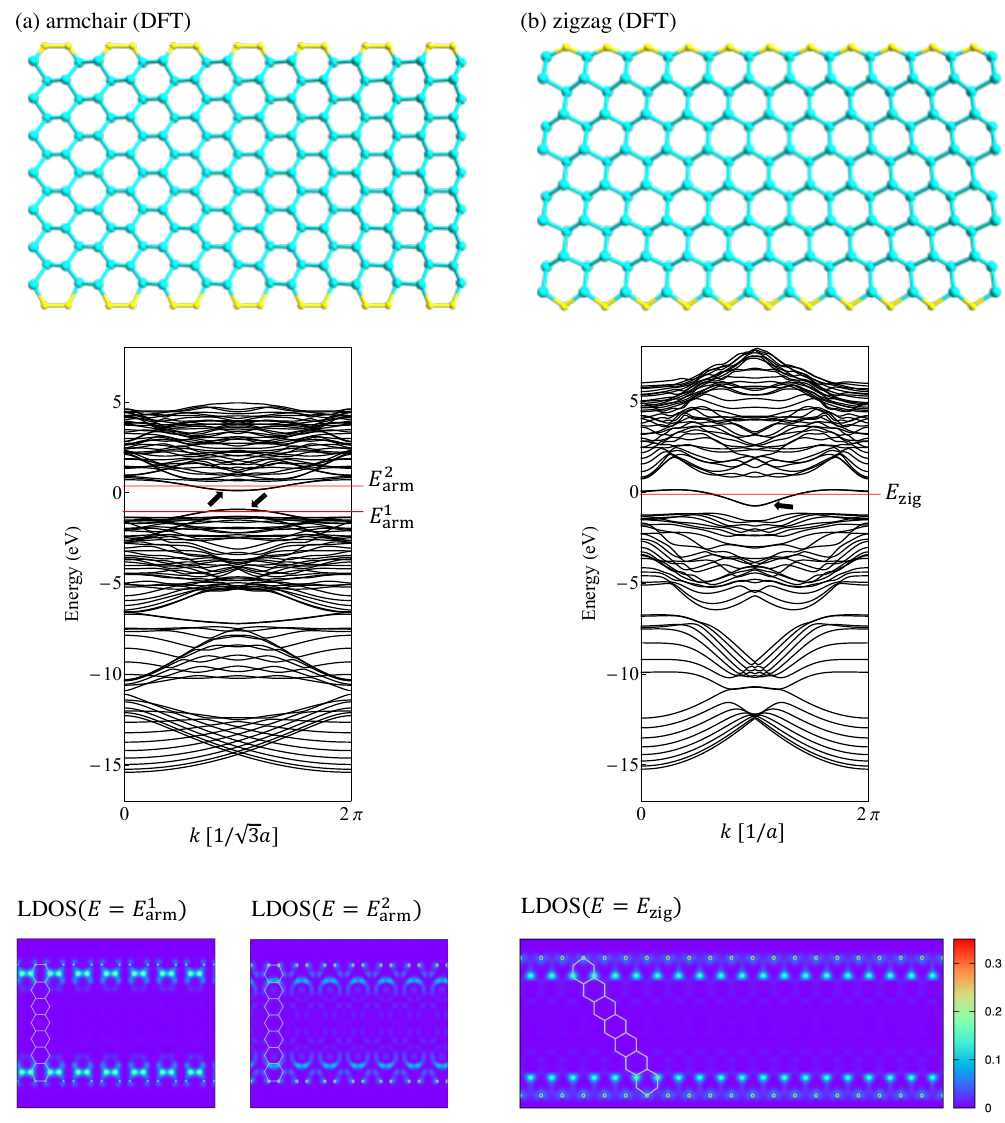}
   \caption{
   (a) The atomic structure (top), the electronic band structure (middle),
   and the local density of states (LDOS, bottom) of a relaxed armchair nanoribbon 
   calculated by DFT.
   In the atomic structure, yellow sites indicate the edge atoms 
   which are relaxed to decrease the buckling height.
   In the band structure, the edge states are marked by arrows, 
   and the red horizontal lines indicate energies $E_{\mathrm{arm}}^{1} = -1.05 \mathrm{eV}$ and $E_{\mathrm{arm}}^{2} = 0.23 \mathrm{eV}$ at which the LDOS is computed in the bottom panel.
   In the LDOS plot, the atomic positions are shown by an overlaid honeycomb lattice.
   (b) Corresponding plots for a relaxed zigzag nanoribbon. 
   The LDOS is calculated at $E_{\mathrm{zig}} = -0.15 \mathrm{eV}$.
    }\label{fig:ribbonstructure_DFTband}
 \end{center}
 \end{figure*}

We calculate the band structures of armchair and zigzag blue phosphorene nanoribbons by using DFT. 
The lattice structures of the ribbons are depicted in Fig.~\ref{fig:ribbonstructure_DFTband}.
The super unit cells of the armchair and zigzag ribbons consist of 30 and 16 atoms, respectively.
We take surface reconstruction into account, where we allow the edge atoms [yellow sites in Fig.~\ref{fig:ribbonstructure_DFTband}] to be relaxed until the forces acting on the nuclei are less than $10^{-4}$ Ry/Bohr. 
By the relaxation of the edge atoms, the buckling height of the edge atoms is decreased [the top panel of Fig.~\ref{fig:ribbonstructure_DFTband}]. 

The resulting electronic band structures are shown in the middle panels of 
Fig.~\ref{fig:ribbonstructure_DFTband}.
The Fermi energies are set to be zero.
In the armchair nanoribbon [Fig.~\ref{fig:ribbonstructure_DFTband}(a)],
we observe a pair of in-gap bands which are repelled from each other, forming a gap at the Fermi energy.
On the other hand, in the zigzag nanoribbon [Fig.~\ref{fig:ribbonstructure_DFTband}(b)], a single band appears in the bulk gap and the Fermi energy is in the middle of the band.
By examining the local density of states (LDOS) of these electronic states [the bottom panel of Fig.~\ref{fig:ribbonstructure_DFTband}], we confirm that these in-gap states are localized around the edges of the ribbon, namely, they are indeed edge states.
 
\subsection{Topological origin of edge states}
\label{subsec:origin of edge}
The origin of the edge states in the zigzag/armchair nanoribbons can be understood by considering the tight-binding model, in a similar manner to the method for the black phosphorene \cite{Hitomi2021}.
Below we introduce the Slater-Koster tight-binding model to qualitatively reproduce the DFT band structure, and a simplified $90^{\circ}$ model where all the bond angles are deformed to $90^{\circ}$.
In the $90^{\circ}$ model, the emergence of the edge states can be easily understood
by considering the position of the Wannier orbitals.
These edge states survive a continuous deformation from the $90^{\circ}$ model to the Slater-Koster model due to the topological equivalence of the two models.

\subsubsection{Slater-Koster tight-binding model}
We construct a TB model using the Slater-Koster parametrization \cite{Slater1954}.
The model is given by
\begin{equation}
    H^{\mathrm{SK}}_{\alpha \beta} (\vb*{k}) = \sum_{\vb*{R}} t_{\alpha \beta} (\vb*{R}) e^{i \vb*{k} \cdot \vb*{R}},
    \label{TBHamiltonian}
\end{equation}
where $\vb*{R} = n_1 \vb*{a}_1 + n_2 \vb*{a}_2$ with $n_1$ and $n_2$ to be integers, and we take $\alpha, \beta = As, Ap_x, Ap_y, Ap_z, Bs, Bp_x, Bp_y, Bp_z$ as the basis. 
The onsite term $t_{\alpha \alpha}(\vb*{R}=0)$ is parameterized by $\varepsilon_s, \varepsilon_{p_x}, \varepsilon_{p_y},
\varepsilon_{p_z}$ for $s, p_x, p_y, p_z$ orbitals, respectively. Due to the symmetry requirement, they do not depend on A and B sublattice, and we have $\varepsilon_{p_x}= \varepsilon_{p_y}$.
The hopping integrals $(\vb*{R}\neq 0)$ are written in the Slater-Koster form as
\begin{align}
    t_{nm} (\vb*{R}) &= e_n e_m (V_{pp\sigma} (R) - V_{pp\pi} (R)) \label{eq:SKparametrization1}
    \\
    t_{nn} (\vb*{R}) &= e_n^2 V_{pp\sigma} (R) + (1-e_n^2) V_{pp\pi} (R) \label{eq:SKparametrization2}
    \\
    t_{sn} (\vb*{R}) &= e_n V_{sp\sigma} (R) \label{eq:SKparametrization3}
    \\
    t_{ss} (\vb*{R}) &= V_{ss\sigma} (R), \label{eq:SKparametrization4}
\end{align}
 where $n, m = x, y, z$ and $e_n = \vb*{R} \cdot \hat{\vb*{n}} /R$ with $\hat{\vb*{n}}$ being the unit vector in $n$ direction.
 We assume that the parameter $V_{ij\alpha}(R)\, (i,j=s,p; \alpha=\sigma,\pi)$ depends on $R$ exponentially,
 \begin{equation}
   V_{ij\alpha} (R) = V_{ij\alpha}^{(0)} e^{-(R-\tau)/r_{0}},
    \label{V(R)}
\end{equation}
where $\tau = [(a/\sqrt{3})^{2} + d^{2}]^{1/2}$ 
is the distance between the nearest-neighbor sites. 
The decay length $r_0 = (a - \tau) / \ln{10}$ is determined in order to let the second nearest-neighbor hopping $V(a)$ be $0.1$ times the nearest-neighbor hopping $V^{(0)}$ \cite{Tsim2020}. 
Note that $V_{ij\alpha} (\tau) = V_{ij\alpha} ^{(0)}$ by definition.

To determine the band parameters $\varepsilon_\alpha$ and $V_{ij\alpha} ^{(0)}$, we obtain the maximally localized Wannier functions and associated tight-binding model by using the Wannier90 package \cite{Pizzi2020}.
The $V_{ij\alpha}^{(0)}$ are determined so as to best reproduce the nearest neighboring hopping integrals in the Wannier90 tight-binding model,
resulting in
\begin{align}
    &V_{pp\sigma}^{(0)} = 3.60 ~\mathrm{eV},\, V_{pp\pi}^{(0)}=-0.90 ~\mathrm{eV} , \nonumber\\
    &V_{sp\sigma}^{(0)}= 2.09 ~\mathrm{eV},\, V_{ss\sigma}^{(0)} = -1.55 ~\mathrm{eV}.
    \label{eq:SKhopping}
\end{align}
The on-site energy can also be taken from the Wannier90 tight-binding model as
\begin{align}
\varepsilon_s = -10.7~{\rm eV},\,
\varepsilon_{p_x} = \varepsilon_{p_y} = -2.01~{\rm eV},\,
\varepsilon_{p_z} = -2.23~{\rm eV}.
\label{eq:SKonsite}
\end{align}

The bulk band structure of the obtained TB model [Fig.~\ref{fig:bulkDFT90TBband}(b)] qualitatively reproduces that of the DFT calculation [Fig.~\ref{fig:bulkDFT90TBband}(a)].

\subsubsection{$90^{\circ}$ model}
\begin{figure*}
\begin{center}
   \includegraphics [width=170mm]{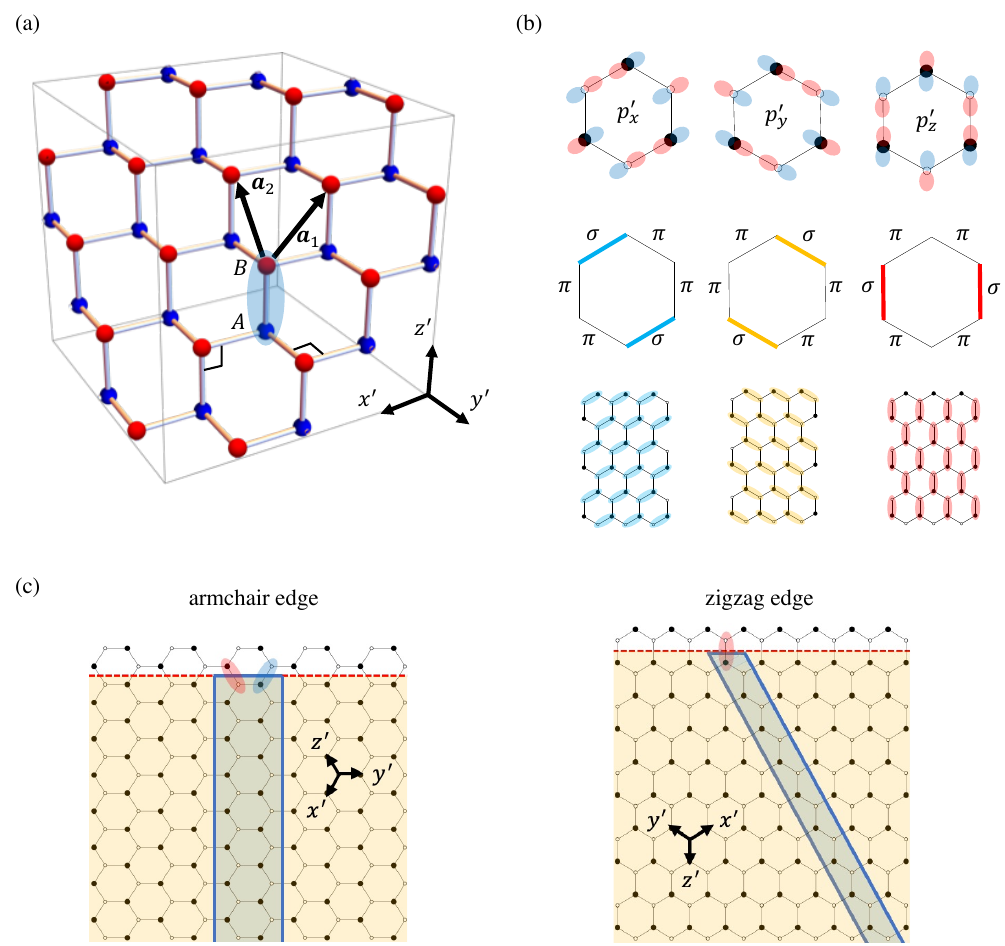}
   \caption{
   (a) The crystal structure of blue phosphorene in the $90^{\circ}$ limit. 
   The blue (red) ball represents the A (B) site.
   (b) Top panel: The schematic figure of the $p_x'$, $p_y'$, $p_z'$ orbitals in the $90^{\circ}$ model. 
   Here, white (black) circle stands for the A (B) site.
   Middle panel: The anisotropic honeycomb lattices corresponding to the $90^{\circ}$ model with only one of the $p_x'$, $p_y'$, $p_z'$ orbitals. 
   The thick bonds indicate the stronger ($\sigma$) bonding. 
   Bottom panel: Three Wannier states which originate from the respective $\sigma$ bonding of the models of the anisotropic honeycomb lattices.
   (c) The half-broken Wannier functions at the edge in the armchair/zigzag nanoribbons. 
   The yellow region represents the ribbon and the blue parallelogram represents the super unit cell. 
   The blue and red Wannier orbitals correspond to $p_x'$ and $p_z'$ orbitals, respectively.
            }\label{fig:90modelexp_new}
 \end{center}
 \end{figure*}

 The $90^{\circ}$ model is a simplified, but topologically equivalent, model of blue phosphorene where all the bond angles are deformed from $93.1^{\circ}$ to $90^{\circ}$ [Fig.~\ref{fig:90modelexp_new}(a)]. 
 By using the model, we can understand the origin of the emergent edge states of blue phosphorene. 
 Here we take the orthogonal $x',y',z'$ axes to be parallel to the bond directions
 and consider only $3s, 3p_{x}', 3p_{y}', 3p_{z}'$ orbitals as the basis.
We incorporate only the nearest-neighbor hopping integrals.
We neglect the hopping between $s$ and $p$ orbitals because the energy bands originating from $s$ orbitals are located far below in energy and the coupling hardly affect the states at the Fermi energy.
The Hamiltoinian of the $90^{\circ}$ model is then written as
\begin{equation}
    H_{90^{\circ}} (\vb*{k}) = \mathrm{diag} [H_s (\vb*{k}), H_{x'} (\vb*{k}), H_{y'} (\vb*{k}), H_{z'} (\vb*{k})],
\label{eq:90Hamiltonian}
\end{equation}  
where the subscripts $s, x', y', z'$ denote the respective $s, p_x', p_y', p_z'$ atomic orbitals.
The $2 \times 2$ matrices $H_{\alpha} (\vb*{k}) (\alpha = s, x', y', z')$ are 
\begin{equation}
    H_\alpha (\vb*{k}) = \mqty(0 & h_{\alpha}(\vb*{k}) \\ h_\alpha(\vb*{k})^{*} & 0) + \varepsilon_{\alpha},
    \label{eq:90Halpha}
\end{equation}
and
\begin{equation}
    h_\alpha (\vb*{k}) = t_{\alpha x'} e^{i \vb*{k} \cdot \vb*{a_1}}
    + t_{\alpha y'} e^{i \vb*{k} \cdot \vb*{a_2}}
    + t_{\alpha z'},
    \label{eq:90halpha}
\end{equation}
\begin{equation}
    t_{\alpha \beta} =  \begin{cases} 
    t_s \ (\alpha = s)\\
    \delta_{\alpha \beta} t_{\sigma} + (1 - \delta_{\alpha \beta}) t_{\pi} \ (\alpha = x', y', z')
    \end{cases}
    \label{eq:talphabeta}
\end{equation}
in the basis of the $A$, $B$ sublattices. 
We take the hopping parameters $t_{\pi} = -1$ eV, $t_{\sigma} = 4$ eV, $t_s = -1.28$ eV, and the on-site energies are $\varepsilon_s = -12$ eV, $\varepsilon_{x'} = \varepsilon_{y'} = \varepsilon_{z'} = 0$. 
These parameters are determined to approximately reproduce the original band structure. Specifically, we apply the Slater-Koster tight-binding model [Eqs. \eqref{eq:SKparametrization1} to \eqref{eq:SKonsite}] to the 90 degree lattice, and round the numbers for simplicity. 
The band structure obtained from Eq.(\ref{eq:90Hamiltonian}) is shown in Fig.~\ref{fig:bulkDFT90TBband}(c).



From Eq.(\ref{eq:90Hamiltonian}), we find that the $s, p_{x}', p_{y}', p_{z}'$ orbitals are completely decoupled,
allowing us to consider each sector individually. 
Focusing on the three $p$ orbitals, each of them is equivalent to a single-orbital TB model on a flat anisotropic honeycomb lattice 
as shown in the middle panel of Fig.~\ref{fig:90modelexp_new}(b). 
For example, we can see the $p_x'$ sector of the $90^{\circ}$ model has stronger $\sigma$-bonds along $x'$ direction
and weaker $\pi$-bonds along $y'$ and $z'$,
and therefore the system is formally equivalent to an anisotropic honeycomb tight-binding model with hopping $t_\sigma$ (thick blue lines) in a single direction
and $t_\pi$ in the other two.
In the anisotropic honeycomb model, it is known that the energy spectrum is gapped when $t_{\sigma} > 2t_{\pi}$  
(it is the case in our $90^{\circ}$ model) \cite{Ezawa2018, Wunsch2008, Pereira2009, Hasegawa2006},
and then the Wannier orbital of the valence band is centered at the midpoint of the strong bond \cite{Ezawa2018, Hitomi2021}. 
By applying the argument to the $90^{\circ}$ model, we immediately see that three Wannier orbitals associated with the $p_x'$, $p_y'$, $p_z'$ orbitals are centered at the inequivalent bond centers, as shown in the bottom panel of Fig.~\ref{fig:90modelexp_new}(b). 

Importantly, the center position of the Wannier orbital
(Wannier center; WC) is a topological invariant, 
i.e. its value is unchanged unless a gap-closing or a symmetry-breaking
occurs \cite{Watanabe2017, Kruthoff2017,  Bradlyn2017,  Song2017,  Zak1982}.
As we will demonstrate in the next section, blue phosphorene and the $90^{\circ}$ model are topologically equivalent, and thus, they both have a WC at each bond center. 
In appendix A, we identify the WCs of blue phosphorene by an alternative method based on the symmetry-based indicator \cite{Watanabe2017, Kruthoff2017,  Bradlyn2017,  Song2017,  Zak1982} and obtain the same result. 

When the WC of occupied bands is mismatched with the atomic positions, the system is classified as an obstructed atomic insulator (OAI).
The blue phosphorene is an OAI because the WCs of the occupied bands 
are centered at the midpoints of the bonds between atomic sites.
In an OAI, edge states appear when the WC is half-broken at the boundary.
The Wannier orbital located at the bond center is nothing but a covalent bond,
and the emergent edge states correspond to a dangling bond (i.e., uncoupled orbitals),
which are energetically isolated from the bulk band region. 
If a WC is cut at the corner of a finite-sized system (e.g., a flake), the corresponding zero-dimensional corner state appears. This is known as a 2D higher-order topological state \cite{Benalcazar2019,Ezawa2018,Hitomi2021, Song2017, Bunney2021, tanay3, tanay6}. 
In the following, we investigate these localized topological states of blue phosphorene 
in detail.

\subsubsection{Origin of the edge states}

We model the armchair and zigzag nanoribbons without surface reconstructions
using the tight-binding models introduced above.
We consider a continuous deformation from the $90^{\circ}$ model to the TB model
\begin{equation}
    H_{\lambda}(k) = (1 - \lambda)H_{90^\circ} (k) + \lambda H_{\mathrm{SK}} (k),
    \label{eq:deformation}
\end{equation}
where $H_{90^\circ} (k)$ and $H_{\mathrm{SK}} (k)$ are
the Hamiltonians of the nanoribbon of the $90^{\circ}$ and the Slater-Koster models, respectively,
and $0 \leq \lambda \leq 1$ is the deformation parameter. 
By diagonalizing $H_{\lambda}(k)$, we obtain the band structure of nanoribbons. 

\begin{figure*}
\begin{center}
   \includegraphics [width=140mm]{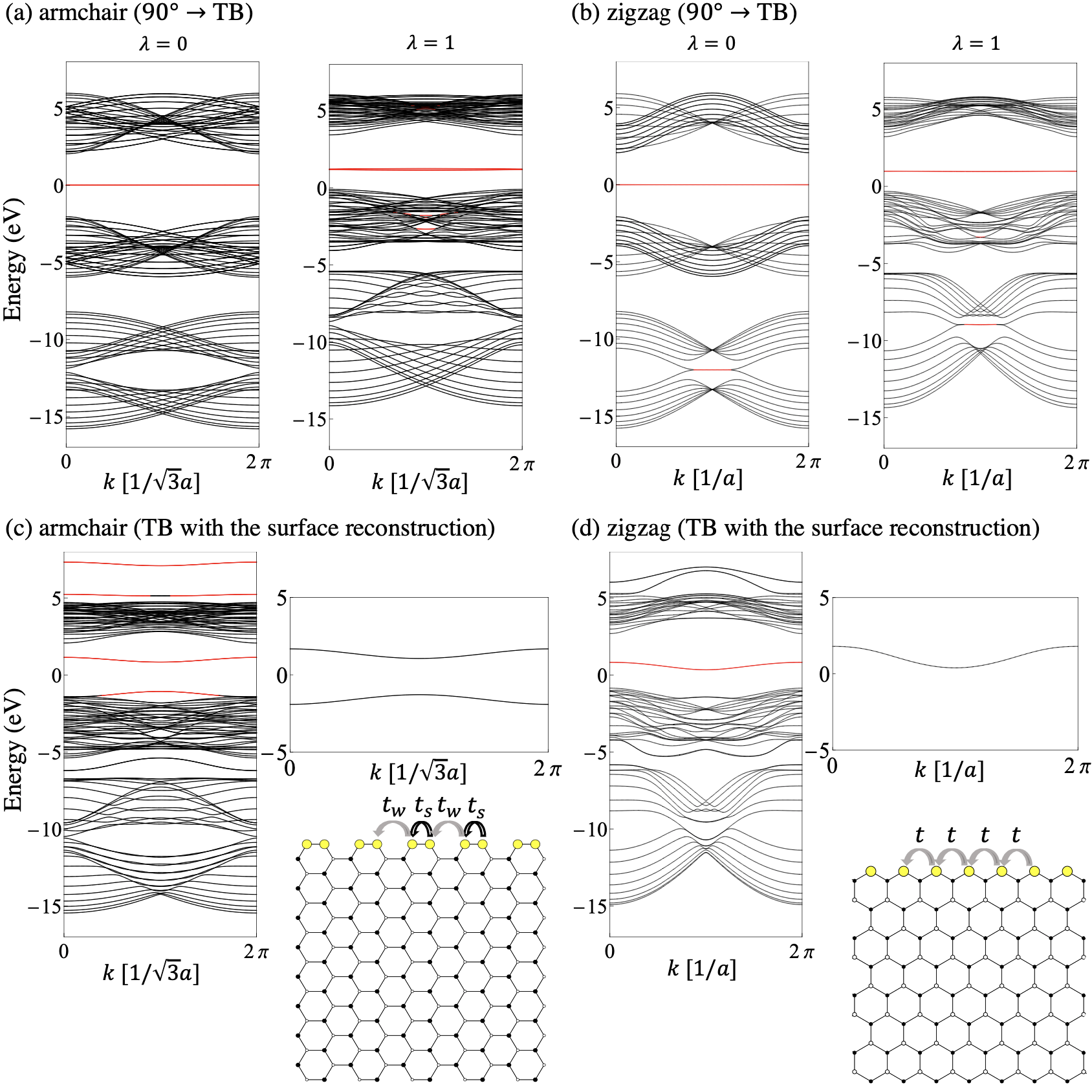}
   \caption{
   (a) Band structure of an armchair nanoribbon in the $90^{\circ}$ model ($\lambda = 0$) and that in the TB model without the surface reconstruction ($\lambda = 1$). 
   The red lines represent the edge-state bands which are two-fold degenerate.
   (b) Corresponding plot for a zigzag nanoribbon. There is only a single edge-state band.
   (c) (Left) Band structure of the armchair nanoribbon in the TB model with the surface reconstruction, where the two edge bands in the main gap are split. (Top right) Energy band of the effective model considering edge uncoupled orbitals,
    which are indicated in the bottom right panel.
   (d) Corresponding plot for the zigzag nanoribbon.
            }\label{fig:deformationandTB}
 \end{center}
 \end{figure*}

For the armchair case, we consider a ribbon with width of $7a$, which is of the same size as the DFT calculation  [Fig.~\ref{fig:ribbonstructure_DFTband}(a)].
Figure \ref{fig:deformationandTB}(a) shows the band structure
of $90^{\circ}$ model ($\lambda = 0$) and that of the Slater-Koster model ($\lambda$ = 1).
The red curves represent the bands of the edge states,
which are defined by the condition that more than $90\%$ of
the probability amplitude is localized within an interval of length $\tau(=a/\sqrt{3})$ 
from the edge sites.
Here we set the cut-off length such that it is around the typical decay length of the in-gap states and much smaller than the ribbon width, to correctly distinguish the bulk states and the edge states.
A small change in the cut-off length does not affect the identification, except for some marginal edge states which are energetically very close to the bulk states.
We observe that the band structure of the Slater-Koster model ($\lambda$ = 1) is deformed into the $90^{\circ}$ model ($\lambda = 0$) without closing the bulk gap, i.e. without a topological phase transition, and hence the WCs shown in the bottom panel of Fig.~\ref{fig:90modelexp_new}(b) are unchanged through the deformation. The edge state bands remain almost intact during the deformation.

The origin of the edge state bands can be explained in terms of broken Wannier orbitals at the boundary of the system.
As argued, the edge states appear when the center of the Wannier orbital
is cut by the boundary. In the case of blue phosphorene, three Wannier functions are localized at the three inequivalent bonds. At the edge of the armchair nanoribbon, two Wannier orbitals corresponding to $p_x'$ and $p_z'$ sectors are broken per super unit cell [Fig.~\ref{fig:90modelexp_new}(c)].
The number of broken Wannier orbitals per super unit cell coincides with the number of the edge state bands per side, $N_{\mathrm{edge}} = 2$.

The same argument can be applied to the zigzag nanoribbon.
Here we consider a ribbon with the super unit cell including 16 atoms,
which is of the same width as the DFT calculation [Fig.~\ref{fig:ribbonstructure_DFTband}(b)].
The band structures under the continuous deformation are shown in Fig.~\ref{fig:deformationandTB}(b).
In this case, we have only one edge-state bands,
corresponding to the fact that the zigzag boundary breaks only 
the Wannier orbital of the $p_z'$ sector in the $90^{\circ}$ limit, as shown in Fig.~\ref{fig:90modelexp_new}(c). 

In the $90^{\circ}$ limit, the edge state bands converge to $E = 0$
both in the armchair and zigzag cases.
This is because the $90^{\circ}$ model has chiral symmetry, i.e.,
the $2\times2$ Hamiltonian for each sector of $\alpha = x',y',z'$ satisfies the condition,
\begin{equation}
 \sigma_z H_{\alpha} (\vb*{k}) \sigma_z = -H_{\alpha} (\vb*{k}) ,
    \label{eq:chiral}
\end{equation}
where $\sigma_z = \mathrm{diag}(1,-1)$. 
Under chiral symmetry, the spectrum is symmetric with respect to $E = 0$.
The two edge states are chiral zero modes satisfying $\sigma_z \psi = \psi$,
which are lock to the zero energy \cite{Ryu2002}.

\subsection{Surface reconstruction}
\label{subsec:recons}

In the Slater-Koster model in the previous section,
we see that the zigzag and armchair edge bands are almost flat, which is the feature inherited from the completely-flat chiral zero modes in the $90^{\circ}$ limit.
In the DFT calculation in Sec.\ \ref{subsec:DFT}, on the other hand, 
the edge bands are much broader in energy, and 
two edge-state bands in the armchair ribbon are repelled away opening an energy gap at the Fermi energy.
Actually, the difference can be explained by incorporating the edge-site reconstruction in the Slater-Koster tight-binding model.


Once we obtain the atomic positions after the surface reconstruction by the DFT calculation [Sec.~\ref{subsec:DFT}], we can immediately construct the tight-binding
Hamiltonian for the relaxed ribbons using the Slater-Koster parametrization introduced in Sec.~\ref{subsec:origin of edge}.
The electronic band structures of the relaxed armchair and zigzag nanoribbons are shown in the left panels of Fig.~\ref{fig:deformationandTB}(c)(d).
We observe that the calculation well reproduces the deformed edge-state bands in the DFT calculation [Fig.~\ref{fig:ribbonstructure_DFTband}(a)(b)].

The relaxed band structure of the edge states can be described by a simple 1D model for the edge atoms.
In the $90^{\circ}$ model, the edge states are contributed by
the uncoupled orbitals at edge sites,
where the edge state bands are completely flat since the hopping integrals between the edge orbitals vanish in the $90^{\circ}$ limit.
By the relaxation, the edge atoms are aligned near the horizontal plane [the top panels of Fig.~\ref{fig:ribbonstructure_DFTband}(a)(b)], and therefore non-zero hopping integrals emerge. 
We can construct an effective 1D Hamiltonian of the edge atoms
by estimating the emergent hopping terms in the following manner.
We write the Slater-Koster Hamiltonian for the relaxed ribbon as
\begin{equation}
    H_{\mathrm{SK, relaxed}} (k) = \mqty(H_{\mathrm{edge}}(k) & U(k) \\ U^{\dag}(k) & H_{\mathrm{bulk}}(k)),
    \label{eq:ribboneff1}
\end{equation}
where $k$ is the 1D wave number, $H_{\mathrm{edge}}$ is the edge sector consisting of the uncoupled orbitals of the edge atoms, $H_{\mathrm{bulk}}$ is the bulk sector composed of the other orbitals and $U$ is the coupling between the edge and bulk sector.
By treating $U$ as a perturbation and projecting $H_{\mathrm{SK, relaxed}}$ onto the edge sector, the effective Hamiltonian is given by
\begin{equation}
    H_{\mathrm{eff}} (k) = H_{\mathrm{edge}}(k) +U^{\dag}(k) \frac{1}{E - H_{\mathrm{bulk}}(k)} U(k),
    \label{eq:ribboneff2}
\end{equation}
where we take $E$ to be the average of the eigenvalues of $H_{\mathrm{edge}}(k)$.
From the effective Hamiltonian $H_{\mathrm{eff}}(k)$, we extract the nearest-neighbor hopping integrals between uncoupled edge orbitals.
For the armchair nanoribbon, $t_s = -1.46$ eV and $t_w = -0.31$ eV, and $t = 0.36$ eV for the zigzag ribbon [the right bottom panel of Fig.~\ref{fig:deformationandTB}(c)(d)].
We can calculate the effective edge band structure analytically, only with the nearest-neighbor hopping integrals (farther hoppings are negligibly small) :
\begin{align}
    &E_{\mathrm{eff},\pm}^{\mathrm{arm}}(k) = \pm[t_{s}^{2} + t_{w}^{2} + 2t_s t_w \mathrm{cos}\sqrt{3}ka]^{1/2}, \\
    &E_{\mathrm{eff}}^{\mathrm{zig}}(k) = 2t\mathrm{cos}ka,
\end{align}
where the respective band widths are $2(t_s + t_w)$ and $4t$. 
For the armchair nanoribbon, the band gap is $2(t_s - t_w)$.
As shown in Fig.~\ref{fig:deformationandTB}(c)(d), the calculated band structures well reproduce those of the original Slater-Koster model. 
\section{Corner States of Blue Phosphorene}
\label{sec:corner}

\subsection{Armchair-armchair corner}
\label{subsec:armcorner}
\begin{figure*}
\begin{center}
   \includegraphics [width=170mm]{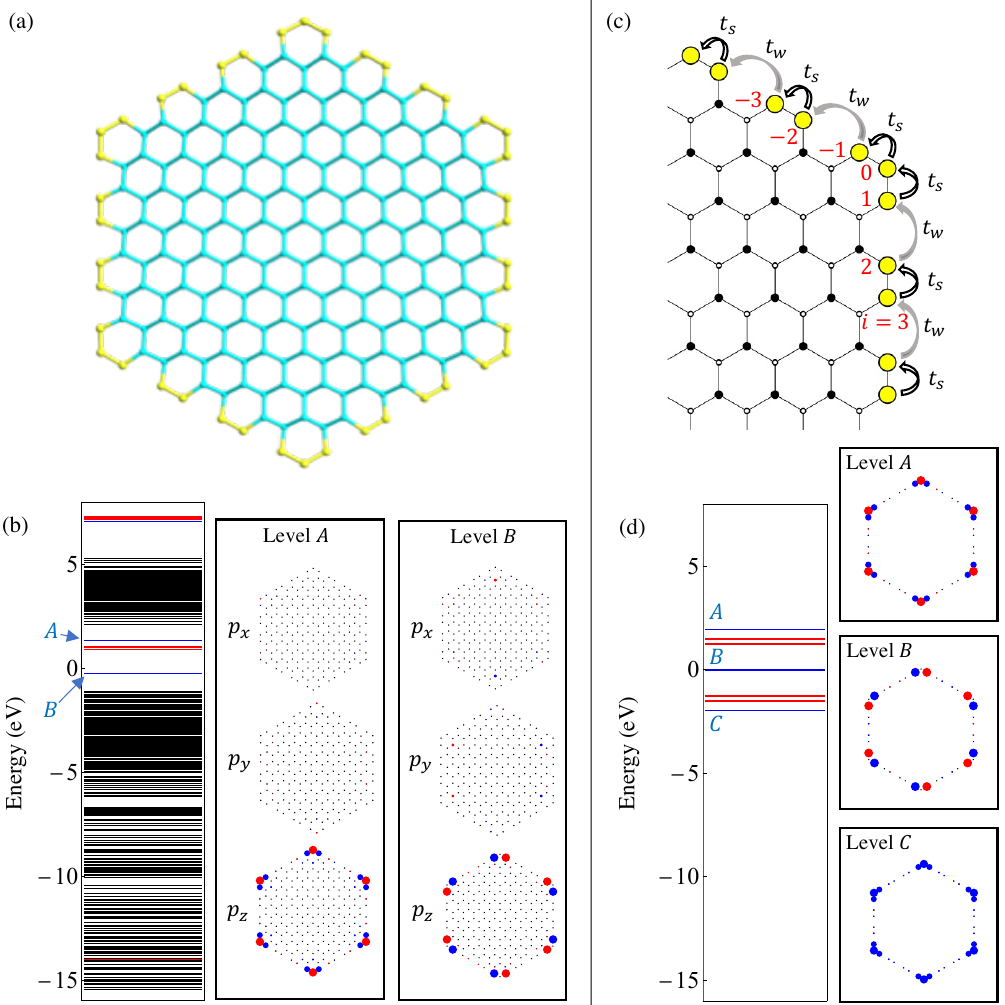}
   \caption{
   (a) The structure of the armchair-armchair flake with the surface reconstruction. 
   Yellow balls are the relaxed edge atoms. 
   The total number of atoms is 222. 
   (b) The energy levels (left) and the wavefunctions (right) of the corner states calculated by the tight-binding model.
   The edge states and the corner states are indicated by red and blue lines, respectively.
   The Fermi energy is in the middle of the B level (see the text).
   In the wavefunction figure, the black dots represent the atomic sites, the radii of colored disks indicate the probability amplitude.
   Red and blue represent positive and negative component in the real part of the wavefunction. 
   (c) The illustration of the effective edge-corner site model.
   (d) The energy levels and the corner state wave functions
   in the effective edge-corner site model.
            }\label{fig:armflake}
 \end{center}
 \end{figure*}

We consider a hexagonal-shaped flake of blue phosphorene
with armchair edges depicted in Fig.~\ref{fig:armflake}(a) (hereafter we call this the armchair flake) by using the 
Slater-Koster TB model constructed in Sec.~\ref{subsec:origin of edge}.
The nanoflake consists of 222 atoms and the surface reconstruction 
of the outermost sites (indicated by yellow atoms) is included
in a similar manner to the nanoribbon.
The whole structure has the crystalline symmetry $D_{3d}$. 
Here we define the corner sites by six atoms at the vertices of the hexagon,
and the edge sites by the rest of the yellow atoms.

We diagonalize the TB Hamiltoinan of the armchair flake to obtain the energy eigenvalues and the eigenstates. 
The calculated energy levels are shown in 
the left panel of Fig.~\ref{fig:armflake}(b),  
where the red and blue lines stand for the edge and corner-localized states, respectively. Here an edge state is defined by
the condition that more than $80\%$ of the total probability amplitude 
is localized at the edge and corner sites, and a corner state is by
that more than $90\%$ of the probability amplitude is localized within a distance of $1.5a$ from the corner sites.
The right panels of Fig.~\ref{fig:armflake}(b) show the wave amplitudes of
three $p$-orbital components for the corner states (the amplitudes of $s$ orbital are negligibly small). 

The bunch of edge levels (red lines) at $E \sim 1$ eV
corresponds to the upper edge-state bands in the armchair nanoribbon [Fig.~\ref{fig:deformationandTB}(c)].
We have two corner levels $A$ and $B$ just above and below the edge levels.
Each of the level A and B is actually composed of six-fold 
degenerate states. 
The degeneracy is slightly broken by weak coupling among the six corners since the system has a finite size,
while they are completely degenerate in the limit of infinite system size. 
The Fermi energy for the charge neutrality is located in the middle of the level B, where three out of six degenerate levels are occupied.

When the Fermi energy is shifted to the gap above the level B (i.e., three electrons are doped to the charge neutral system), a fractional electric charge $-e/2$ appears at each corner point.
This is because three excessive electrons must be equally distributed
to the six corners due to $S_3$ symmetry.
When the Fermi energy is in the gap below the level B, likewise, $+e/2$ appears at each corner.
This situation is so-called \textit{filling anomaly} \cite{Benalcazar2019,Takahashi2021}.
The emergence of corner states in a blue-phosphorene armchair flake
was also reported in very recent work~\cite{Qian2021},
where the corner state is located off the charge neutrality point
since the structural relaxation is not included in the calculation.

Just similar to the case of the armchair ribbon,
these edge and corner states can be well described by 
an effective model only taking the uncoupled $p_z$ orbitals at the boundary.
The Hamiltonian can be obtained by applying Eq.~(\ref{eq:ribboneff2})
to the flake.
Figure \ref{fig:armflake}(c) illustrates the schematic view of the model, 
where the strong hopping ($t_s \approx -1.36$eV) and the weak hopping ($t_w \approx -0.25$eV) are arranged alternately, except that
$t_s$ appears successively at the corner site ($i = 0$).
This simple model yields three corner levels A, B and C
as well as the edge levels between them,
as shown in Fig.~\ref{fig:armflake}(d). 
The wavefunction of the corner states are essentially the eigenstates of a three-site system composed of $i = -1,0,1$.
We see that the energy spectrum and wave functions coincide with the results in the TB model, while the corner level C and the lower edge bands are 
absorbed into the bulk states in the TB model.
The filling anomaly argued above is best understood by this model.
The charge neutral point corresponds to the half filling of the 
six-fold level B, which is the zero mode of the three-site system at each corner.



\subsection{Zigzag-zigzag corner}
\begin{figure*}
\begin{center}
   \includegraphics [width=170mm]{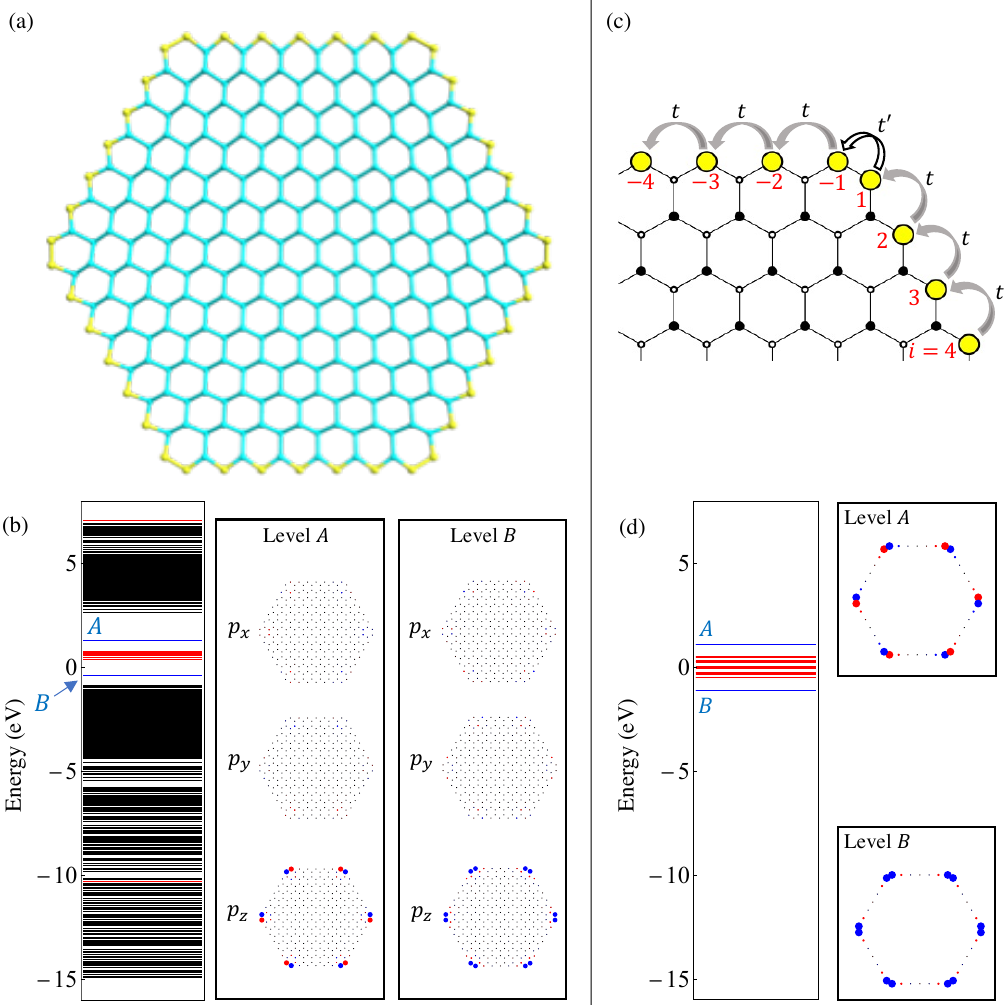}
   \caption{
   (a) The structure of the zigzag-zigzag flake with the surface reconstruction. 
   Yellow balls are the relaxed edge atoms. 
   The total number of atoms is 294.
   (b) The energy level and the wavefunctions of the corner states calculated by the tight-binding model are shown in the same manner as Fig.~\ref{fig:armflake}.
   (c) The illustration of the effective edge site model.
   (d) The energy level and the corner states of the effective edge site model.
            }\label{fig:zigflake}
 \end{center}
 \end{figure*}

We also investigate a nanoflake with zigzag edges as shown in Fig.~\ref{fig:zigflake}(a). 
The system also has $D_{3d}$ symmetry. 
We obtain the eigenenergies and the eigenstates using the TB model 
with the relaxed edge structure.
The result is summarized in Fig.~\ref{fig:zigflake}(b). 
We have the two corner levels A and B which are both six-fold degenerate. 
Unlike the armchair flake, the Fermi energy for the charge neutral point is located in the middle of the edge bands between the two corner states,
and hence we do not have isolated fractional corner charge when the Fermi energy is
shifted.

The effective edge-site model derived by using the same method
is dipicted in Fig.~\ref{fig:zigflake}(c).
There are no corner sites ($i = 0$), and $i = 1$ and $-1$
are directly connected. 
The nearest-neighbor hopping parameter between $i$ and $i+1$
takes a constant value $t = 0.29$eV,
while the bond between $i = 1$ and $-1$ is $t' = -1.03$eV.
The model qualitatively reproduces the spectrum 
and the characters of the two corner states as shown in Fig.~\ref{fig:zigflake}(d). 
Obviously, the level A and B can be understood as 
the antibonding and bonding states, respectively, 
at the irregular bond between $i = 1$ and $-1$.


\section{Comparison with Black Phosphorene}
\label{sec_discussion}

\begin{figure}
\begin{center}
   \includegraphics [width=85mm]{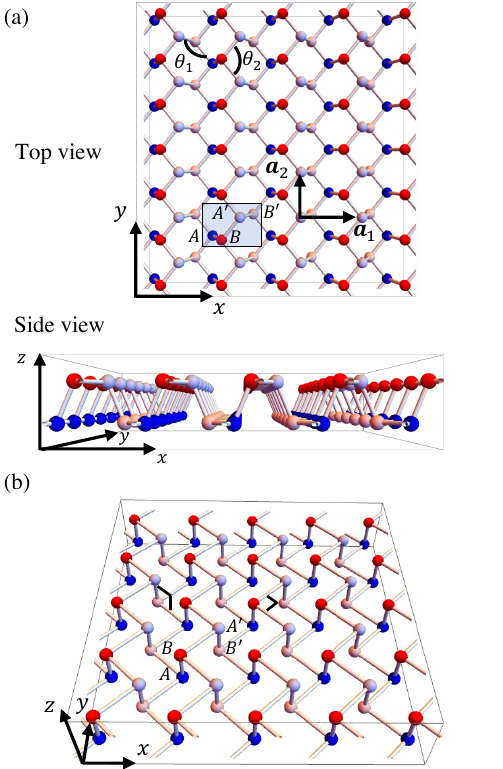}
   \caption{
   (a) The crystal structure of black phosphorene from the top and the side views. 
   The bond angles are $\theta_{1} \sim 103^{\circ}$ and $\theta_{2} \sim 98^{\circ}$.
   The unit cell is shown as the blue rectangular, with A, B, A', B' sublattices. 
   (b) The $90^{\circ}$ model of black phosphorene. 
   Both $\theta_1$ and $\theta_2$ are deformed to $90^{\circ}$.
            }\label{fig:lattice_blackP}
 \end{center}
 \end{figure}

The topological nature of blue phosphorene
is analogous to that of its close cousin, black phosphorene~\cite{Yu2015,Guo2014,Khandelwal2017,Hitomi2021}.
The crystal structure of black phosphorene is shown in Fig.~\ref{fig:lattice_blackP}(a).
Locally it has a non-flat three-bonded structure similar to that of blue phosphorene,
while the buckling directions at $A$ and at $A'$ are opposite, 
resulting in a completely different global structure.
The blue phosphorene belongs to the symmorphic space group $P\bar{3}m1$, and black phosphorene to the nonsymmorphic space group $Pmna$. 
Because of the nonsymmorphic structure,
a unit cell of black phosphorene consists of four atoms,
while blue phosphorene consists of two.

It is notable that, in spite of these differences, the Hamiltonian of black phosphorene is equivalent to that of blue phosphorene
within the $90^{\circ}$ model with the nearest neighbor hopping,
as explained in the following.
Fig.~\ref{fig:lattice_blackP}(b) illustrates the $90^{\circ}$ model for black phosphorene. 
Here we see that a vertical bond from $A$ to $B$
and one from $A'$ to $B'$ are opposite i.e., in $+z$ and $-z$ directions, respectively.
If we reverse the bond $A'B'$ from $-z$ to $+z$, then
the entire structure becomes identical to the $90^{\circ}$ model for blue phosphorene  [Fig.~\ref{fig:90modelexp_new}(a)].
In this process, however, the Hamiltonian matrix is not modified
because the hopping integral between two $p_z$ orbitals
aligned along $z$ direction does not change if we swap the 
positions of the two orbitals,
and the hopping among $p_x$ and $p_y$ orbitals are all zero
for vertical bonds.
Therefore, the $p_xp_yp_z$ sector of $90^{\circ}$-model Hamiltonian of black phosphorene is identical to that of the blue phosphorene.




This immediately concludes that 
the blue and black phosphorenes share the same Wannier orbital centers, and thus
the same properties in emergence of the edge/corner states~\cite{Hitomi2021}.
Indeed, it was previously shown 
that the zigzag edge of  black phosphorene has a single edge-state band, 
and the armchair edge has two edge-state bands~\cite{Fukuoka2015, Guo2014, Hitomi2021}.
Also, the first principle calculation for black phosphorene \cite{Guo2014} showed 
that two edge-state bands of an armchair edge
are split by the lattice relaxation, in a similar manner to our result
for blue phosphorene.
We expect that the filling anomaly in an armchair-armchair corner 
also occurs in black phosphorene, too,
because it is owing to splitting of the armchair-edge bands
such that a corner state can come to the charge neutral point.




\section{Conclusions}
\label{sec_conclusion}
We have examined the edge and corner states of blue phosphorene 
and investigated their origins in relation to the center positions of Wannier functions.
We found that the existence of the Wannier orbitals at every bond center yields edge states both in zigzag and armchair edges. 
The band structures and wavefunctions of the edge states
can be described by a simple effective Hamiltonian  
for uncoupled boundary orbitals. 
In particular, the effective model qualitatively explains 
the effect of the surface relaxation on the band structure
in the first-principles calculation.


We investigated two types of nanoflakes consisting of zigzag/armchair edges,
and found that several corner-localized states emerge in the bulk gap.
These modes are again explained by a similar effective model considering
edge and corner uncoupled orbitals.
In the armchair flake, particularly, 
we demonstrated that corner states appear right at the Fermi energy,
leading to the emergence of fractional corner charge. 

Finally, we discussed the relationship between blue phosphorene
and black phosphorene.
Although the two systems have completely different atomic structures,
we showed that they share the equivalent Wannier orbital positions 
and similar edge/corner state properties.


\begin{acknowledgements}
This work was supported by JSPS KAKENHI Grants No. JP21J20403, No. JP20K14415, No. JP20H01840, No. JP20H00127, No. JP21H05236, No. JP21H05232, and by JST CREST Grant No. JPMJCR20T3, Japan. 
\end{acknowledgements}

\appendix
\section{Wannier center of blue phosphorene and symmetry-based indicator}
As shown in the main text, positions of Wannier centers are crucial to understanding 
the topological origin of edge and corner states in the blue phosphrene. 
In Sec.~\ref{subsec:origin of edge}, these positions are identified in terms of continuous deformation between the blue phosphorene and the $90^{\circ}$ model.
On the other hand, there is another generic scheme to detect the Wannier centers from the symmetry character of occupied bands~\cite{Watanabe2017,Kruthoff2017,Bradlyn2017,Song2017,Zak1982}. 
In this Appendix, we show that the latter scheme also gives the consistent results with Sec.~\ref{subsec:origin of edge}.


Quite generally, each single particle states are classified into the irreducible representations (irreps) at the high-symmetry points in $\vb*{k}$-space.
We list these irreps for space group $P\bar{3}m1$ specific to the blue phosphorene in Table~\ref{tab:irrep}.
Here, the high-symmetry points are $\Gamma=(0,0)$, $K=(4\pi/3a)(1, 0)$ and $M=(2\pi/\sqrt{3}a)(0, 1)$.
The irreps allow us to describe the symmetry character of a set $S$ of bands by a vector called symmetry-based indicator,
\begin{equation}
    \vb*{b}_S = (\gamma_{1}^{+}, \gamma_{2}^{+}, \gamma_{3}^{+}, \gamma_{1}^{-}, \gamma_{2}^{-}, \gamma_{3}^{-} ; \kappa_1, \kappa_2, \kappa_3 ; \mu_{1g}, \mu_{2g}, \mu_{1u}, \mu_{2u}),
    \label{eq:irreps}
\end{equation}
where $\gamma_{i}^{\pm}$, $\kappa_i$ and $\mu_{is}$ ($i = 1, 2, ...$ and $s = u, g$) are the number of irreps $\Gamma_{i}^{\pm}$, $K_i$ and $M_{is}$, respectively, for the bands included in the set $S$. 



   \begin{table} [hbtp]
       \centering
        \caption{
        The irreducible representations (irreps) for the three high-symmetry points are given.
        We show the character of each irrep rather than the representation matrix.
        $\Gamma_{i} {}^{\pm}$, $K_i$, $M_{is}$ are the irreps at $\Gamma$, $K$, $M$ points, respectively. 
        $C_{2x}$ is the $180^{\circ}$ rotation along the $x$ axis, $P$ is the spatial inversion and $\sigma_{x}$ is the mirror reflection about the $y$-$z$ plane. The little group of $D_{3d}$ (crystallographic point group of blue phosphorene) at $K$ point is $D_3$ which do not contain $P$ and $\sigma_{x}$ operations, so the rows of them for $K$ point are blank.}
        \begin{tabularx}{85mm}{XXXX}
    \hline \hline 
        Irrep & $C_{2x}$ & $P$ & $\sigma_{x}$ \\ [1mm]
    \hline
        $\Gamma_{1}^{+}$ & $+1$ & $+1$ & $+1$ \\ [1mm] 
         $\Gamma_{1}^{-}$ & $+1$ & $-1$ & $-1$ \\ [1mm] 
          $\Gamma_{2}^{+}$ & $-1$ & $+1$ & $-1$ \\ [1mm] 
           $\Gamma_{2}^{-}$ & $-1$ & $-1$ & $+1$ \\ [1mm] 
            $\Gamma_{3}^{+}$ & $0$ & $+2$ & $0$ \\ [1mm] 
             $\Gamma_{3}^{-}$ & $0$ & $-2$ & $0$ \\ [1mm] 
             $K_1$ & $+1$ & --- & --- \\ [1mm] 
             $K_2$ & $-1$ & --- & --- \\ [1mm] 
             $K_3$ & $0$ & --- & --- \\ [1mm] 
             $M_{1g}$ & $+1$ & $+1$ & $+1$ \\ [1mm] 
             $M_{1u}$ & $+1$ & $-1$ & $-1$ \\ [1mm] 
             $M_{2g}$ & $-1$ & $+1$ & $-1$ \\ [1mm] 
             $M_{2u}$ & $-1$ & $-1$ & $+1$ \\ [1mm] 
    \hline \hline
    \end{tabularx}
       \label{tab:irrep}
   \end{table}

\begin{figure}
 \begin{center}
   \includegraphics [width=85mm]{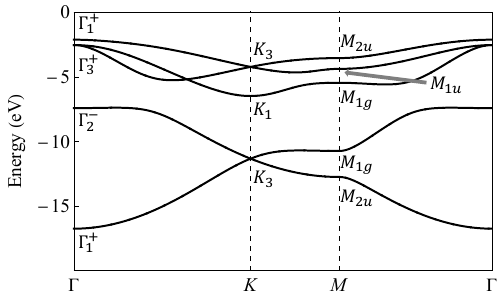}
   \caption{The irreducible representations of the occupied bands ($E < 0$). $3p$ and $3s$ orbitals are included.
            }\label{fig:irreps}
 \end{center}
 \end{figure}

Let us identify the indicator~(\ref{eq:irreps}) of occupied bands in the blue phosphorene. 
We specify the irreps of the occupied bands in Fig.~\ref{fig:irreps}, which is obtained by considering the symmetry of the wave function of the each single particle states.
The indicator corresponding to the irreps in Fig.~\ref{fig:irreps} is
\begin{equation}
    \vb*{b}_{\mathrm{occ}} = (2, 0, 2, 0, 1, 0 ; 1, 0, 4 ; 2, 0, 1, 2).
\end{equation}


Not all components of the indicator (\ref{eq:irreps}) are independent, but they are connected by compatibility relations,
\begin{gather}
    \gamma_{1}^{+} + \gamma_{1}^{-} + \gamma_{2}^{+} + \gamma_{2}^{-} +  \gamma_{3}^{+} + \gamma_{3}^{-} = M, \label{eq:c1} \\
    \kappa_1 +\kappa_2 + \kappa_3 = M, \label{eq:c2} \\
    \mu_{1g} + \mu_{1u} + \mu_{2g} + \mu_{2u} = M, \label{eq:c3} \\
    \gamma_{1}^{+} + \gamma_{1}^{-} + (\gamma_{3}^{+} + \gamma_{3}^{-})/2 = \kappa_1 +\kappa_{3}/2, \label{eq:c4} \\
    \gamma_{1}^{+} + \gamma_{2}^{-} + (\gamma_{3}^{+} + \gamma_{3}^{-})/2 = \mu_{1g} + \mu_{2u} \label{eq:c5}.
\end{gather}
Here, Eqs.~(\ref{eq:c1})-(\ref{eq:c3}) guarantee that the number of valence bands is fixed to $M$ at every $\vb*{k}$-point. 
The conditions (\ref{eq:c4}) and (\ref{eq:c5}) forbid the band crossing of opposite-parity states under the $180^{\circ}$ rotation $C_{2x}$ and the reflection $\sigma_x$, respectively. 
These compatibility relations (\ref{eq:c1})-(\ref{eq:c5}) reduce the 13 degrees of freedom of the indicator (\ref{eq:irreps}) to 9 components:
 \begin{equation}
    \tilde{\vb*{b}}_S = (\gamma_{1}^{+}, \gamma_{2}^{+}, \gamma_{3}^{+}, \gamma_{1}^{-}, \gamma_{2}^{-} ; \kappa_1 ; \mu_{1g}, \mu_{2g} ; M).
    \label{eq:irrepstilde}
\end{equation}
In this representation, the indicator for the occupied bands in blue phosphorene is written as
\begin{equation}
    \tilde{\vb*{b}}_{\mathrm{occ}} = 
    (2, 0, 2, 0, 1 ; 1 ; 2, 0 ; 5).
    \label{eq:irrepstilde}
\end{equation}

 \begin{figure}
 \begin{center}
   \includegraphics [width=85mm]{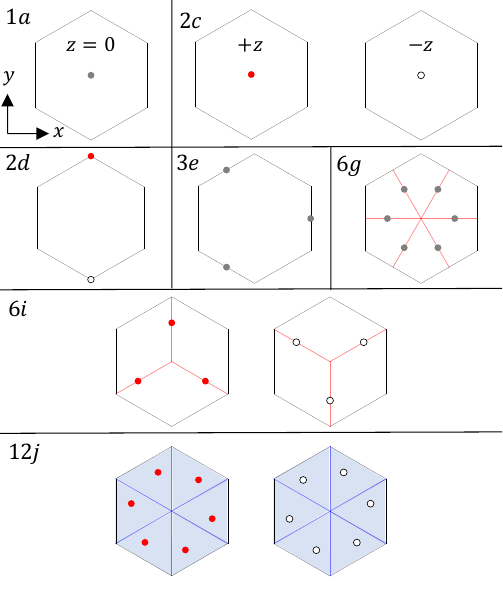}
   \caption{
   The Wyckoff positions of the space group $P\bar{3}m1$. 
   Each of the vertices of the hexagon corresponds to the atomic site of blue phosphorene. 
   Gray, red and white dots represent the positions with the heights $0$, $z$ and $-z$, respectively.
   Positions $6g$ and $6i$ can move on the red thick lines. 
   $12j$ is the general positions and each dot can be located at any position in the blue triangular region.
            }\label{fig:WP}
 \end{center}
 \end{figure}

To detect the Wannier centers of blue phosphorene from the indicator~(\ref{eq:irrepstilde}), 
we use the concept of elementary band representations~\cite{Bradlyn2017}.
The elementary band is a set of band structure obtained from possible atomic orbitals, which are restricted from the symmetry.
Given the symmetry of the system, we can classify the positions in the unit cells into Wyckoff positions (WPs).
Each of the WPs are characterized by its site-symmetry group (SSGs),
namely the space subgroup keeping the WP invariant up to the lattice vectors.
Irreps of the SSG give the possible atomic orbitals (e.g., $s$-like, $p_z$-like) at the WP.
We summarize all of the WPs for the space group $P\bar{3}m1$ in Fig.~\ref{fig:WP} and all SSGs for the WPs in the second column of Table \ref{tab:WPirrep}.
For example, the SSG of WP $3e$ is the point group $C_{2h}$, which consists of the $180^{\circ}$ rotation along $x$-axis ($C_{2x}$) and the mirror reflection about $y$-$z$ plane ($\sigma_x$), 
see Fig.~\ref{fig:WP}.
The possible orbitals for this WP $3e$ are $A_g$ ($s$-like), $A_u$ ($p_x$-like), $B_g$ ($d_{xz}$-like), and $B_u$ ($p_{y}$-like), see also Table~\ref{tab:WPirrep}.
In the same way as Eq.(\ref{eq:irrepstilde}) the symmetry-based indicator also characterize the elementary bands generated from each of atomic orbitals:
\begin{equation}
\begin{split}
    \tilde{\vb*{b}}_{\chi} = (\gamma_{1}^{+}{}^{(\chi)}, \gamma_{2}^{+}{}^{(\chi)}, \gamma_{3}^{+}{}^{(\chi)}, \gamma_{1}^{-}{}^{(\chi)}, \gamma_{2}^{-}{}^{(\chi)} 
    ; \kappa_{1}^{(\chi)} ; \\
    \mu_{1g}^{(\chi)}, \mu_{2g}^{(\chi)} ; M^{(\chi)}),
    \end{split}
    \label{eq:chiindicator}
\end{equation}
which are listed in the 5th-13th columns of Table \ref{tab:WPirrep}. Here $\chi=1,2,\cdots, 21$ is a serial number of the atomic orbitals.


 \begin{table*} [hbtp]
       \centering
        \caption{
        The symmetry-based indicators of the 21 elementary bands for the space group $P\bar{3}m1$. 
        All of the WPs are listed in the 1st column and the corresponding SSGs are shown in the 2nd column. 
        For these SSGs, several irreps which designate the symmetries of atomic orbitals are displayed in the 3rd column. 
        For each of them labeled by the serial numbers $\chi$, we show the indicator (\ref{eq:chiindicator}) in the 5th-13th columns. 
        }
        \begin{tabularx}{170mm}{XXXXXXXXXXXXX}
    \hline \hline 
        WP & SSG & irreps & $\chi$ & $\gamma_{1}^{+}$ & $\gamma_{2}^{+}$ &  $\gamma_{3}^{+}$ & $\gamma_{1}^{-}$ & $\gamma_{2}^{-}$ & $\kappa_1$ & $\mu_{1g}$ & $\mu_{2g}$ & $M$ \\ [1mm]
    \hline
    $1a$ & $D_{3d}$ & $A_{1g}$ & 1 & 1 & 0 & 0 & 0 & 0 & 1 & 1 & 0 & 1 \\ [1mm]
     &  & $A_{1u}$ & 2 & 0 & 0 & 0 & 1 & 0 & 1 & 0 & 1 & 1 \\ [1mm]
     &  & $A_{2g}$ & 3 & 0 & 1 & 0 & 0 & 0 & 0 & 0 & 0 & 1 \\ [1mm]
     &  & $A_{2u}$ & 4 & 0 & 0 & 0 & 0 & 1 & 0 & 0 & 0 & 1 \\ [1mm]
     &  & $E_{g}$ & 5 & 0 & 0 & 2 & 0 & 0 & 0 & 1 & 0 & 2 \\ [1mm]
     &  & $E_{u}$ & 6 & 0 & 0 & 0 & 0 & 0 & 0 & 0 & 1 & 2 \\ [1mm]
     $2c$ & $C_{3v}$ & $A_{1}$ & 7 & 1 & 0 & 0 & 0 & 1 & 1 & 1 & 0 & 2 \\ [1mm]
     &  & $A_{2}$ & 8 & 0 & 1 & 0 & 1 & 0 & 1 & 0 & 1 & 2 \\ [1mm]
     &  & $E$ & 9 & 0 & 0 & 2 & 0 & 0 & 0 & 1 & 1 & 4 \\ [1mm]
     $2d$ & $C_{3v}$ & $A_{1}$ & 10 & 1 & 0 & 0 & 0 & 1 & 0 & 1 & 0 & 2 \\ [1mm]
     &  & $A_{2}$ & 11 & 0 & 1 & 0 & 1 & 0 & 0 & 0 & 1 & 2 \\ [1mm]
     &  & $E$ & 12 & 0 & 0 & 2 & 0 & 0 & 1 & 1 & 1 & 4 \\ [1mm]
     $3e$ & $C_{2h}$ & $A_{g}$ & 13 & 1 & 0 & 2 & 0 & 0 & 1 & 1 & 1 & 3 \\ [1mm]
     &  & $A_{u}$ & 14 & 0 & 0 & 0 & 2 & 1 & 2 & 1 & 1 & 3 \\ [1mm]
     &  & $B_g$ & 15 & 1 & 2 & 0 & 0 & 0 & 1 & 0 & 1 & 3 \\ [1mm]
     &  & $B_u$ & 16 & 0 & 0 & 0 & 1 & 2 & 1 & 1 & 0 & 3 \\ [1mm]
      $6g$ & $C_{2}$ & $A$ & 17 & 1 & 0 & 2 & 1 & 0 & 2 & 2 & 2 & 6 \\ [1mm]
     &  & $B$ & 18 & 0 & 1 & 2 & 0 & 1 & 0 & 1 & 1 & 6 \\ [1mm]
     $6i$ & $C_{s}$ & $A'$ & 19 & 1 & 0 & 2 & 0 & 1 & 1 & 2 & 1 & 6 \\ [1mm]
     &  & $A''$ & 20 & 0 & 1 & 2 & 1 & 0 & 1 & 1 & 2 & 6 \\ [1mm]
      $12j$ & $C_{1}$ & $A$ & 21 & 1 & 1 & 4 & 1 & 1 & 2 & 3 & 3 & 12 \\ [1mm]
    \hline \hline
    \end{tabularx}
       \label{tab:WPirrep}
   \end{table*}

In terms of the indicator, any Wannier representable band structure is decomposed into the linear combination of the elementary bands:
\begin{equation}
    \tilde{\vb*{b}}_{S} = \sum_{\chi = 1}^{21} n_{\chi} \tilde{\vb*{b}}_{\chi}.
\end{equation}
For the occupied bands in blue phosphorene, the decomposition is uniquely determined as 
\begin{equation}
   \vb*{b}_{\mathrm{occ}} = \vb*{b}_{10} + \vb*{b}_{13}.
   \label{eq:BluePdecomposition}
\end{equation}
Here $\vb*{b}_{10}$ represents the $s$ orbital of the phosphorus atom located at the WP $2d$.
The remaining $\vb*{b}_{13}$ is the contribution from $p_x$, $p_y$, $p_z$ orbitals of the phosphorus.
According to Table~\ref{tab:WPirrep}, they construct three $s$-like orbitals centered at $3e$ corresponding to the bond centers between phosphorus atom, 
which is the same as the Wannier center obtained in the main text.


\bibliography{reference}
\end{document}